\documentclass[review,10pt]{elsarticle}
\usepackage{lineno,hyperref}
\modulolinenumbers[5]
\usepackage{amsmath}
\usepackage{amsfonts}
\usepackage{amssymb}
\usepackage{lmodern}
\usepackage{xfrac}
\usepackage{graphicx}
\graphicspath{ {./figures/} }
\usepackage{subfig}
\usepackage{epsfig}
\usepackage{epstopdf}
\usepackage{textcomp}
\usepackage{multirow}
\usepackage{setspace}
\usepackage{fullpage}
\usepackage{float}
\usepackage{caption}
\usepackage{siunitx}
\usepackage{xcolor}
\usepackage{tabularx, booktabs}
\usepackage{enumitem}
\usepackage{dcolumn}% Align table columns on decimal point
\usepackage{bm}% bold math
\usepackage[T1]{fontenc}
\biboptions{sort&compress}

\begin{document}
\begin{frontmatter}

% Use the \preprint command to place your local institutional report number 
% on the title page in preprint mode.
% Multiple \preprint commands are allowed.
%\preprint{}

\title{\large{Time series forecasting of multiphase microstructure evolution using deep learning}}

%% Group authors per affiliation:
\cortext[contrib]{These authors contributed equally.}
\cortext[mycorrespondingauthor]{Corresponding author.}
%\cortext[cor1]{\mbox{Corresponding author. \textit{Email address:} \texttt{gsupriyo2004@gmail.com}} (Supriyo Ghosh)}
\author{Saurabh Tiwari\corref{contrib}}
\author{Prathamesh Satpute\corref{contrib}}
\author{Supriyo Ghosh\corref{mycorrespondingauthor}}
\ead{supriyo.ghosh@mt.iitr.ac.in; gsupriyo2004@gmail.com}
\address{Department of Metallurgical and Materials Engineering, Indian Institute of Technology, Roorkee, UK 247667, India}

%\author[mymainaddress]{\corref{mycorrespondingauthor}}
%\cortext[mycorrespondingauthor]{Corresponding author}
%\ead{}

%% or include affiliations in footnotes:
%\author{}
%\ead[url]{www.elsevier.com}

%\author[mymainaddress]{\corref{mycorrespondingauthor}}
%\cortext[mycorrespondingauthor]{Corresponding author}
%\ead{}
%

\date{\today}

\begin{abstract}
Microstructure evolution, which plays a critical role in determining materials properties, is commonly simulated by the high-fidelity but computationally expensive phase-field method. To address this, we approximate microstructure evolution as a time series forecasting problem within the domain of deep learning. Our approach involves implementing a cost-effective surrogate model that accurately predicts the spatiotemporal evolution of microstructures, taking an example of spinodal decomposition in binary and ternary mixtures. Our surrogate model combines a convolutional autoencoder to reduce the dimensional representation of these microstructures with convolutional recurrent neural networks to forecast their temporal evolution. We use different variants of recurrent neural networks to compare their efficacy in developing surrogate models for phase-field predictions. On average, our deep learning framework demonstrates excellent accuracy and speedup relative to the ``ground truth'' phase-field simulations. We use quantitative measures to demonstrate how surrogate model predictions can effectively replace the phase-field timesteps without compromising accuracy in predicting the long-term evolution trajectory. Additionally, by emulating a transfer learning approach, our framework performs satisfactorily in predicting new microstructures resulting from alloy composition and physics unknown to the model. Therefore, our approach offers a useful data-driven alternative and accelerator to the materials microstructure simulation workflow.
\end{abstract}


\end{frontmatter}

\section{Introduction}\label{sec_intro}
Microstructure evolution during materials processing determines the macroscopic properties of the material. Understanding and predicting the time history of microstructure evolution is therefore essential. The phase-field (PF) model is the most high-fidelity method for simulating moving boundary problems such as microstructure evolution~\cite{tourret2022_review,tonks2019_review,steinbach2009,moelans2008}. This method uses a diffuse interface representation, which obviates the need for explicit interface tracking. The microstructure evolution relies on the solution of a coupled system of nonlinear partial differential equations (PDEs) involving continuous field variables (\textit{e.g.}, composition, phase, temperature) that evolve both in space and time, making the PF model computationally intensive. Several approaches are employed in the literature to accelerate the PF simulations, including high-performance parallel computing and advanced numerical schemes such as adaptive mesh refinement and nonlinear solution approaches to multiphysics applications~\cite{ghosh2022tusas,sakane2022parallel,boccardo2023}. However, the associated computational cost remains high, especially for (i) interfacial dynamics problems, where resolving intricate spatial length scales can be essential as in additive manufacturing~\cite{ghosh2023_review}; (ii) large-scale long-time simulations~\cite{steinmetz2016large,stephen2022}; (iii) simulation of multi-component multi-phase material systems~\cite{steinmetz2016large,ghosh20183d}; and (iv) when a large number of simulations are required to understand parameter-microstructure relationships~\cite{huang2024predictive}.

Deep learning (DL) models can provide a viable cost-minimal alternative to microstructure simulation (for recent reviews, see~\cite{fan2021_review,wikle2023statistical}). A DL model encompasses multiple connected machine learning components to learn trends in simulation data involving strong dependencies in space (\textit{e.g.}, images) and time (\textit{e.g.}, sequences) without prior knowledge of the underlying physical mechanisms. These advantages lead to the recent ``exploding'' interest in applications of DL in materials research. Deep neural models, such as convolutional neural network (CNN) and recurrent neural network (RNN), have architectures particularly suited for the given problem, noting that microstructural evolution is essentially a time series problem. An additional challenge, however, remains particularly for handling a large number of model parameters and input/output time frames, affecting the efficiency and speedup of the DL model. To address this, we use a convolutional autoencoder to reduce the dimensional description of the input microstructures in the latent space. These ``compressed'' microstructures subsequently train the convolutional recurrent neural network (CRNN) to predict future microstructure evolution in the autoencoder latent space. The CRNN framework leverages CNN models to extract the spatial features of the microstructure efficiently and utilizes RNN to forecast microstructure evolution. We are especially interested in RNNs due to their internal ``memory'' units, which enable them to ``remember'' microstructure evolution history required for time-series forecasting. We compare RNNs such as simple RNN, long short-term memory (LSTM), and gated recurrent unit (GRU) to gauge their efficacy in developing surrogate models for phase-field predictions. In particular, the LSTM and GRU networks make use of memory cells and gating mechanisms to overcome the vanishing and exploding gradient problems that arise during backpropagation through time (BPTT) in the simple RNN model, enabling them to capture long-term temporal dependencies in the datasets. The GRU offers a simpler alternative to LSTM with fewer tensor operations, allowing for faster training. For a comprehensive tutorial overview of these deep models, refer to~\cite{fan2021_review,geron2022hands}. Once the CRNN architecture makes a future prediction, the decoder segment in the autoencoder takes the predicted microstructure in the latent space and reconstructs it back to the original dimensions. The unified approach of convolutional autoencoder and RNN makes an efficient surrogate model -- a simplified model designed to represent and substitute a computationally expensive, high-fidelity phase-field model.

Deep learning of microstructure evolution described by PDE-based phase-field simulations remains a challenging topic. Several data-driven approaches exist in the literature for learning and predicting spatiotemporal data generated by PF simulations of binary spinodal decomposition and grain microstructure evolution. These studies utilize combination of LSTM and autoencoder architecture~\cite{oommen2022learning,ahmad2023accelerating}, combination of two-point spatial correlations, principal component analysis (PCA), and LSTM~\cite{montes2021accelerating}, Eidetic 3D LSTM (by applying convolution to both the spatial and the temporal dimensions)~\cite{yang2021self}, predictive RNN (by modifying the memory cells to accommodate spatiotemporal flow)~\cite{farizhandi2023spatiotemporal}, combination of different dimensional reduction methods and RNN models~\cite{hu2022accelerating}, tensor-decomposition approaches~\cite{wu2023emulating,iquebal2023emulating}, combination of CNN and transfer learning~\cite{farizhandi2022deep}, and convolutional artificial neural networks~\cite{peivaste2022machine}. 

The majority of the above studies focused on a single deep model with ``simple'' two-phase material microstructures in which both phases possess the same morphology, which minimizes the inherent variance and randomness during microstructure evolution compared to a three-phase system, where different types of morphology can coexist within the same microstructure~\cite{huang1995phase,amoabeng2017,balluffi2005kinetics}. We use different variants of RNN to compare their performance by forecasting the evolution of two-phase and three-phase microstructures. We generate the evolution datasets from PF simulations of spinodal decomposition in binary and ternary systems widely found in measurements of metal alloys and polymer blends~\cite{balluffi2005kinetics,Porter}. The resulting microstructure evolution involves complex curvature-driven interface migration and long-range nonlinear diffusion of multiple chemical species as described by the phase-field equations~\cite{ghosh_pccp,ghosh2020_jcp,ghosh2024_fractal}. To learn a low-dimensional representation of such data on a linear manifold, for example by PCA, may lead to additional spatiotemporal information loss compared to nonlinear embedding techniques such as the autoencoder. Therefore, in this work, we report an autoencoder-based RNN framework as a microstructure learning engine, which can then efficiently be used for the time series forecasting of microstructure evolution without the need to solve the computationally-expensive phase-field governing equations.  

The paper is organized as follows. In Sec.~\ref{sec_pf}, we discuss the phase-field models for microstructure evolution. In Sec.~\ref{sec_ml_methods}, we briefly describe the components of the deep model employed for phase-field predictions. In Sec.~\ref{sec_results}, we present the phase-field results as the true microstructures and compare them with deep model predictions to illustrate the performance of our surrogate model. We discuss the future potential enhancements of deep learning approaches as a microstructure evolution emulator in Sec.~\ref{sec_discussion}. We summarize and conclude in Sec.~\ref{sec_summary}.

\section{Phase-field simulations of microstructure evolution: Ground truth}\label{sec_pf}
For training and testing of the surrogate model, we specifically focus on two-dimensional simulations of spinodal decomposition in binary and ternary alloy systems. We use Cahn-Hilliard-based phase-field simulations that have been detailed elsewhere~\cite{ghosh_pccp,ghosh2020_jcp,Abi}. We generate two datasets from these simulations. The first dataset describes a binary mixture ($A_{1/2}B_{1/2}$) phase separating into two-phase microstructures. The second dataset represents spinodal decomposition in a ternary alloy ($A_{1/4}B_{1/4}C_{1/2}$), yielding three-phase microstructures. These typical compositions represent the common spinodal morphologies, including bicontinuous and isolated networks of phases within a matrix phase~\cite{Puri_review,amoabeng2017}.   

We simulate the spinodal decomposition of the binary alloy $A_{1/2}B_{1/2}$ by solving the equation~\cite{jokisaari2017benchmark}:
\begin{equation}\label{eq_binary}
\frac{\partial c_A (r, t)}{\partial t} =  \nabla \cdot M \nabla \left[ 2c_A(1-c_A)(1-2c_A) - \kappa \nabla^2 c_A\right],
\end{equation}
where $c_A$ is the local volume fraction of component $A$ as a function of position $r$ and time $t$ obeying $c_A + c_B = 1$, $M$ is the mobility, and $\kappa$ is the gradient energy coefficient. 

We simulate spinodal decomposition of the ternary alloy $A_{1/4}B_{1/4}C_{1/2}$ by simultaneously solving the following equations~\cite{ghosh_pccp,ghosh2020_jcp}:
\begin{equation}\label{eq_ternary}
\begin{aligned}
\frac{\partial c_A(r, t)}{\partial t} &= M_{AA}\left[\nabla^2 \lbrace\ln c_A-\ln c_C+\left(\chi_{AB}-\chi_{BC}\right)c_B+\chi_{AC}\left(c_C-c_A\right)\rbrace -2(\kappa_A + \kappa_C)\nabla^4 c_A - 2\kappa_C\nabla^4 c_B\right] \\
&-M_{AB}\left[\nabla^2 \lbrace\ln c_B-\ln c_C+\left(\chi_{AB}-\chi_{AC}\right)c_A+\chi_{BC}\left(c_C-c_B\right)\rbrace -2(\kappa_B + \kappa_C)\nabla^4 c_B-2\kappa_C\nabla^4 c_A\right],  \\
\frac{\partial c_B (r, t)}{\partial t} &= M_{BB}\left[\nabla^2 \lbrace\ln c_B-\ln c_C+\left(\chi_{AB}-\chi_{AC}\right)c_A+\chi_{BC}\left(c_C-c_B\right)\rbrace-2(\kappa_B + \kappa_C)\nabla^4 c_B - 2\kappa_C\nabla^4 c_A\right] \\
&-M_{AB}\left[\nabla^2 \lbrace\ln c_A-\ln c_C+\left(\chi_{AB}-\chi_{BC}\right)c_B+\chi_{AC}\left(c_C-c_A\right)\rbrace-2(\kappa_A + \kappa_C)\nabla^4 c_A-2\kappa_C\nabla^4 c_B\right].
\end{aligned}
\end{equation}
Here, $c_i$ is the local volume/mole fraction of component $i$ such that $c_A + c_B + c_C = 1$. The effective mobility parameters are given by: $M_{AA} = \left(1-c_A\right)^2 M_A + c_A^2 \left(M_B + M_C\right)$, $M_{BB} = \left(1-c_B\right)^2 M_B + c_B^2 \left(M_A + M_C\right)$, and $M_{AB} = M_{BA} = \left(1-c_A\right) c_B M_A + c_A \left(1-c_B\right) M_B - c_A c_B M_C$, with $M_A$, $M_B$, $M_C$ are the mobility of respective components. The $\chi_{ij}$ is the effective interaction energy between components $i$ and $j$, approximated using the regular solution model~\cite{ghosh2020_jcp}. The $\kappa_{i}$ is the gradient energy coefficient of $i$.

We primarily use a 128 $\times$ 128 domain with a uniform grid spacing of $\Delta x = \Delta y = 1.0$, timestep size of $\Delta t = 0.005$, and periodic boundary conditions in all directions. In Eq.~\eqref{eq_binary}, $\kappa = 1$ and $M = 1.0$. In Eq.~\eqref{eq_ternary}, $M_{AA} = M_{BB} = 1.0$ and $M_{AB} = 0.5$, $\chi_{AB} = \chi_{BC} = \chi_{AC} = 3.5$, and $\kappa_A = \kappa_B = \kappa_C = 4.0$. We add random noise of variance $\pm$ 0.01 to the initial configuration to initiate phase separation. We implement with the semi-implicit Fourier spectral method for numerical integration.

We use an in-house phase-field code~\cite{ghosh_pccp,ghosh2020_jcp} that generates the spatiotemporal microstructure dataset by solving Eqs.~\eqref{eq_binary} and~\eqref{eq_ternary}. We collect the microstructure sequences generated after phase-field simulations to construct the dataset for the surrogate model. Each simulation produced 10000 images. The system becomes phase-separated after 500 frames for the ternary mixture and thus is used as a reference time to compare between binary and ternary systems. The time interval between subsequent frames corresponds to 200 timesteps ($\Delta t$ = 0.005). 

%Training and optimization of autoencoder and CRNN architechtures
\section{Surrogate model architecture: Accelerated phase-field framework} \label{sec_ml_methods}
\begin{figure}[htbp]
\centering
\includegraphics[width=\textwidth]{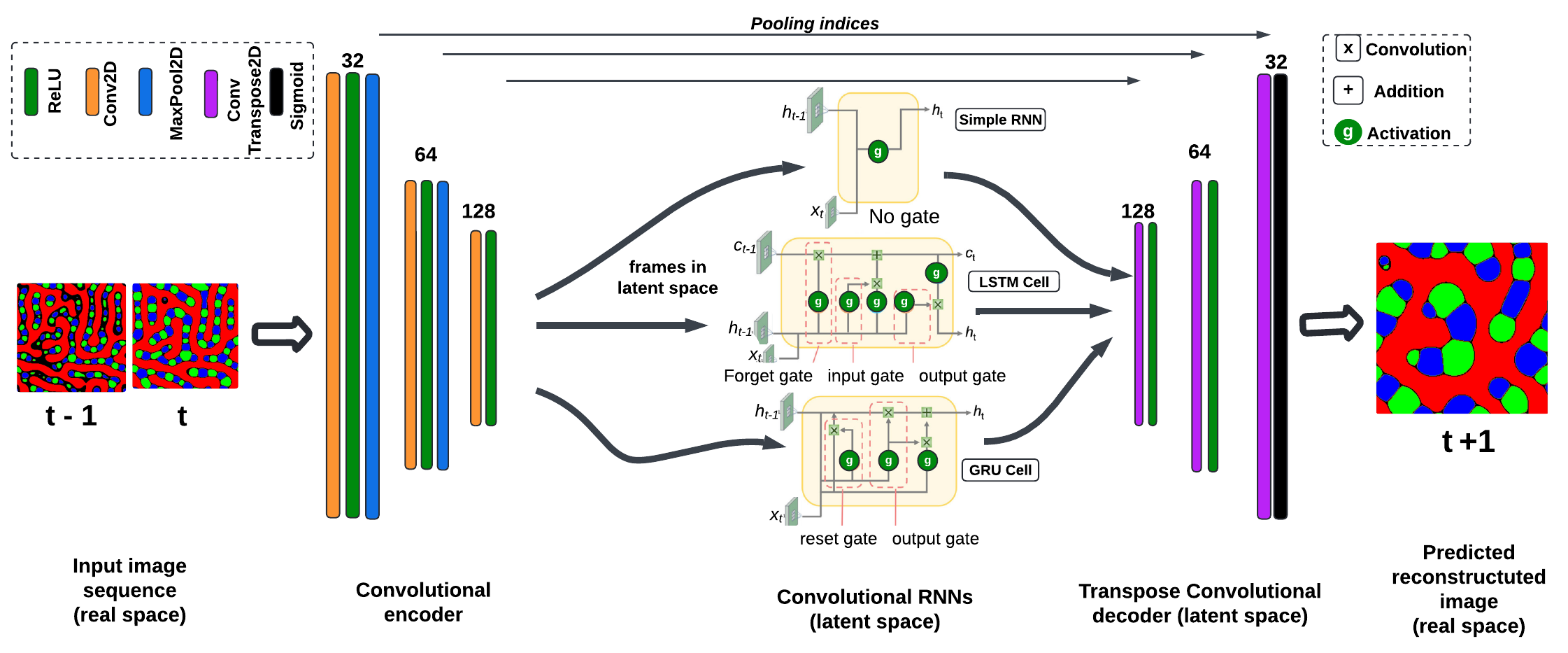}
\caption{(Color online) The schematic represents the workflow of our data-driven approach, which combines the convolutional autoencoder and convolutional recurrent neural network (RNN) containing one of the RNN models from simple RNN, long short-term memory (LSTM), and gated recurrent unit (GRU). The $x_t$, $h_t$, and $c_t$ are the input, hidden, and cell states at current time $t$ (and $t-1$ is prior timestep). The form of the activation function ($g$) varies depending on the gating operations in LSTM and GRU networks. Our workflow predicts the next frame ($t+1$) in sequence based on the input of two consecutive frames (\textit{i.e.}, $t-1$ and $t$) generated from the phase-field model. Although we have shown one RNN cell for reference, two cells are used to process the first input of 2-frame clips ($t-1$ and $t$) to predict frame $t+1$.}\label{fig_schematic}
\end{figure}

As mentioned in Sec.~\ref{sec_intro}, our data-driven approach combines a convolutional autoencoder (encoder-decoder network) with the RNN model. We use simple RNN, LSTM, and GRU as the RNN models. A schematic of our workflow is given in Fig.~\ref{fig_schematic}.

The convolutional layers in CNN process 40000 color images in total, with the size of the largest frame used for experiments being 512 $\times$ 512 $\times$ 3 (red, green, and blue make three channels). Thus, we compress the frames to reduce the computational load, the memory usage, and the number of model parameters. The encoder reduces the dimensional representation of the input frames in the latent space that are subsequently fed to the RNN model to predict the output frame (still in the latent space) that the decoder reconstructs back to the real space. The proposed CNN encoder-decoder architecture is illustrated in Fig.~\ref{fig_schematic}. We construct a typical CNN architecture by stacking three 2D convolutional layers (alias: Conv2D) in the encoder, with each layer consisting of a given number of filters (initially 32, then 64, then 128), kernel size of 3 $\times$ 3, the rectified linear unit (ReLU) activation, and a max-pooling layer (alias: MaxPool2D, kernel size of 2 $\times$ 2, stride = 2, and padding = 1). The decoder includes three transpose convolutional layers (alias: ConvTranspose2D). The last output layer uses the sigmoid activation function.

The output from the encoder consists of a time series representing the microstructure frames in latent space $l_d$ with $d = 32$. The RNN model takes a sequence of two image inputs (say, frames $t - i$ and $t$) as a \textit{training data point} with $i$ being the interval between frames and, for each input sequence, it outputs a single value: the forecast for timestep $t+i$. We use $i = 10$ throughout the study. We use the same hyperparameters for all three RNNs to compare their performance in predicting target microstructures. In these experiments, we set epochs = 25, batch size = 4, number to cells = 2, learning rate = 0.0005, and the dropout rate and recurrent dropout defaults to 0. We set the CNN kernel to be 3 $\times$ 3 and the number of channels of each hidden state to be 128.

Since phase separation begins after $t = 500$ frames of microstructure evolution, we split the frames between $t \in \lbrace500,\ 5000\rbrace$ between the training set $t \in \lbrace500,\ 900\rbrace$ and the testing set $t \in \lbrace900,\ 5000\rbrace$. Although we could have used a much larger training data size to further improve the RNN's accuracy, this smaller size was good enough to capture the satisfactory performance of the model. Moreover, we use mini-batch learning: a batch of two-frame sequence with 38 total sequences between frames $t \in \lbrace500,\ 900\rbrace$ are used to train the workflow. We train the model three times before using it to make predictions. We use two types of representative tests. First, we ask RNN to forecast the next frame in sequence for the given two-frame input sequence from the test dataset. Second, for the same two-frame input, we ask RNN to forecast multiple frames in series, \textit{e.g.}, 20 future frames equivalent to 200 timesteps. For example, with the input frames of $t = 900$ and $t = 910$, we ask RNN to predict the future frames in time series from $t \in \lbrace920,\ 1210\rbrace$ at an interval of $t = 10$. We test the quantitative accuracy of each future prediction by feeding them as an input configuration to the phase-field solver for obtaining the correct long-term microstructure evolution trajectory.

We use the TensorFlow~\cite{tensorflow} framework for implementing our data model. We use the standard He~\cite{he2015delving} initialization to initialize the convolutional kernels (including weights and biases). We use the Adam~\cite{adam_optimizer} optimizer to optimize the framework. Unless otherwise noted, we use default configurations as accessible by TensorFlow~\cite{tensorflow} variables, including the hyperparameter values for both CNN and RNNs. We use the Kaggle~\cite{kaggle} workspace to implement the surrogate model in Python and run it directly in its notebook environment using graphics processing unit (GPU) accelerator (NVIDIA T4 $\times$ 2 configuration). In order to evaluate the model performance, we use the mean squared error (MSE) as the loss function, which evaluates the error between the ground truth and predictions by~\cite{james2013_statistics}:
\begin{equation}\label{eq_mse}
MSE = \frac{1}{N} \sum_{i}^{N} (Y_i - \hat{Y}_i),
\end{equation}
where $Y_i$ is the pixel-wise predicted output, $\hat{Y}_i$ is the pixel-wise value of the ground truth, and $N$ is the total number of pixel in each frame. 

\section{Results and discussion}\label{sec_results}
\subsection{Simulation of ground truth and data preparation for surrogate model}
\begin{figure}[htbp]
\centering
\includegraphics[width=\textwidth]{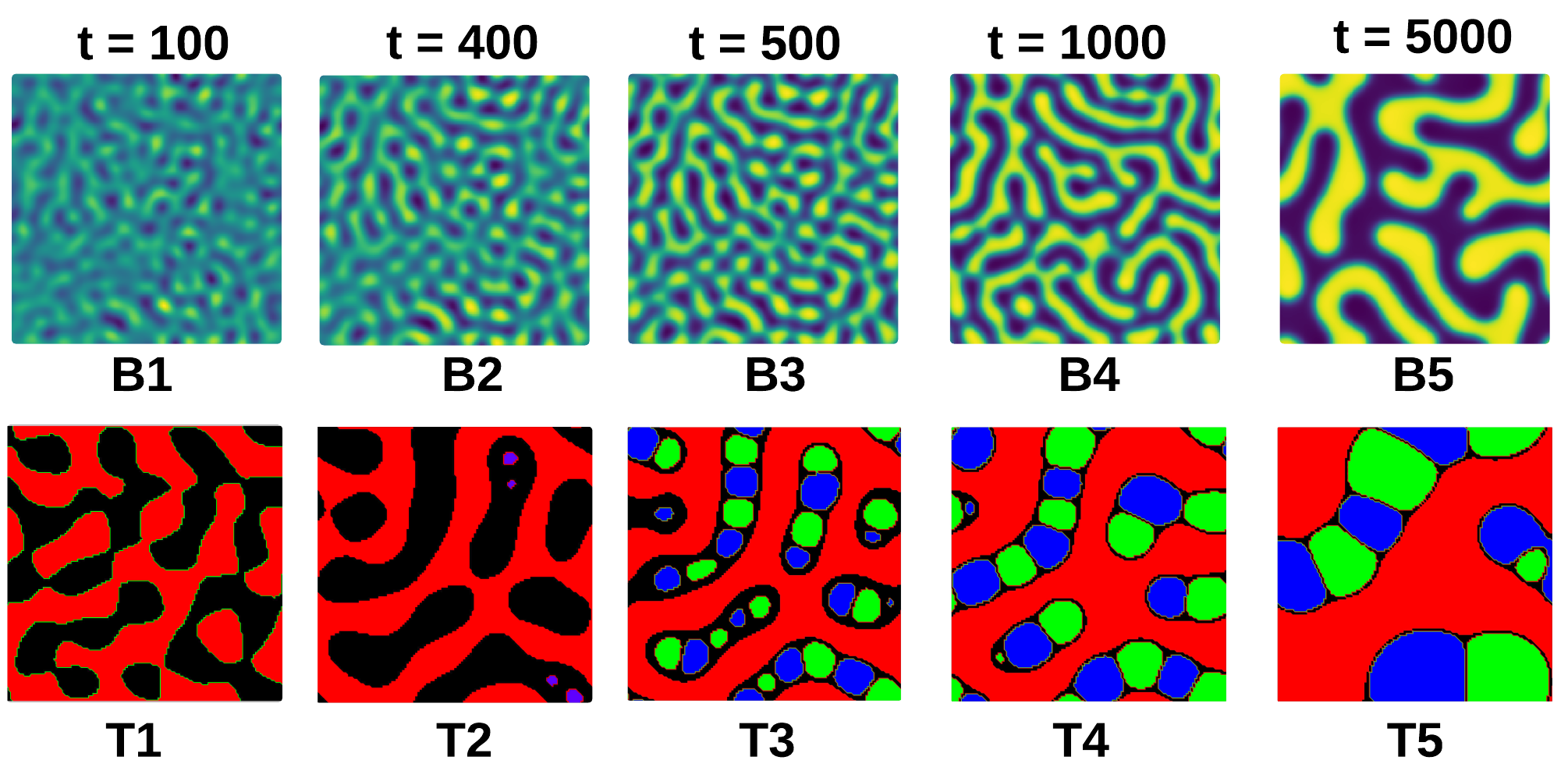}
\caption{(Color online) Phase-field simulation results at timestep $t$ is shown. (Top row: B1-B5) Spinodal decomposition in a binary mixture leads to two-phase microstructure evolution (blue: $A$ and yellow: $B$). (Bottom row: T1-T5) Spinodal decomposition in a ternary mixture leads to three-phase microstructure evolution (green: $A$, blue: $B$, and red: $C$). Spinodal decomposition is followed by coarsening of the phases over an extended period ($t = 5000$ and beyond). Since all the phases in the ternary system form around $t = 500$, our data-driven study begins from this frame.}\label{fig_pf_images}
\end{figure}

We use phase-field simulations to generate two-phase and three-phase microstructure datasets (Sec.~\ref{sec_pf}). Each simulation produced 10000 consecutive images. We show the temporal microstructure evolution process in Fig.~\ref{fig_pf_images}. To minimize the total free energy of the system, small random compositional fluctuations that we add to the initial mixture ($t = 0$) amplify over time, leading to phase-separation of the mixture at an earlier time ($t = 500$) followed by coarsening (Ostwald ripening~\cite{voorhees1985}) of the phases in the intermediate ($t = 1000$) to late times ($t = 5000$). The microstructure evolution pathways of two-phase and three-phase microstructures are drastically different. Hence, they are used as inputs to the surrogate model to demonstrate the model performance. The two-phase mixture typically shows an interconnected bi-continuous network of the phases, while the three-phase microstructure shows the dispersion of two isolated globular phases within a continuous domain of the third phase. These microstructure evolution images are used as the ground truth to train the surrogate model for forecasting time series of microstructure evolution. Spinodal decomposition typically proceeds \textit{via} rapid formation of phases followed by slow, steady evolution of the phases with the characteristic length scale gradually increases over time due to the Gibbs-Thomson effect~\cite{Porter}. For the ternary system, the rapid composition modulation happens over a short time span, $t = 450$ to $t = 500$. Hence, we train our data model since frame $t = 500$ when microstructure evolution enters a steady state regime. Our goal is to develop an accelerated phase-field framework with limited training data to examine the RNN's extrapolating capability for forecasting microstructure evolution multiple times ahead of the last training data. We train the surrogate model using time frames in $t \in \lbrace500,\ 900\rbrace$. Thus, the frames in $t \in \lbrace900,\ 5000\rbrace$ become the test data.

\begin{figure}[htbp]
\centering
\subfloat[Binary system]{\label{a}\includegraphics[width=0.4\textwidth]{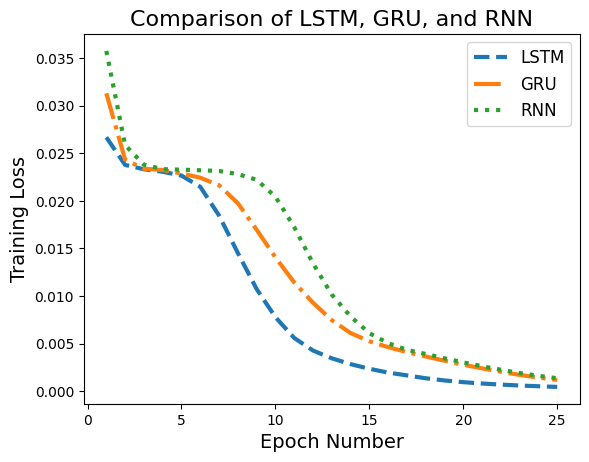}}
\subfloat[Ternary system]{\label{b}\includegraphics[width=0.4\textwidth]{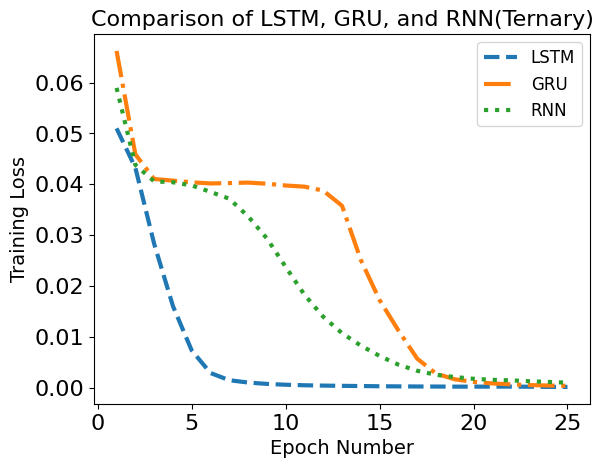}}
\caption{The value of the loss function (\textit{i.e.}, MSE) obtained during model training is shown as a function of epoch number (or iterations) for experiments with (a) two-phase and (b) three-phase microstructure evolution. The horizontal plateau close to the bottom of the plot indicates convergence after a given number of iterations. The LSTM converges faster than both simple RNN and GRU.}\label{fig_loss}
\end{figure}

Our framework is based on low-dimensional representation of the microstructure spatial statistics that evolve over time. As illustrated in Fig.~\ref{fig_schematic}, we input a sequence of microstructure frames to the autoencoder to reduce their dimensional representation. The 2D convolutional layers in the encoder ``squeeze'' the input images in the latent space. The resulting image sequences are fed to the convolutional RNN for future predictions. Since our focus is to perform a \textit{comparative} study on the predictive ability of the different RNN models, we use a very basic cell structure with two RNN cells to process the input of two consecutive frames with a frame interval of 10, leading to a total number of 2-frame training points to 38. This optimal setup allows computational affordability yet fast convergence during model training, as shown in Fig.~\ref{fig_loss}. The learning curves as a function of the number of epochs further indicate that LSTM converges faster than simple RNN and GRU. The output prediction from the RNN model propagates through the decoder, which reconstructs it back to the original dimensions.

\subsection{One-frame ahead forecasting by RNNs}\label{sec_next}
\begin{figure}[htbp]
\centering
\includegraphics[width=\textwidth]{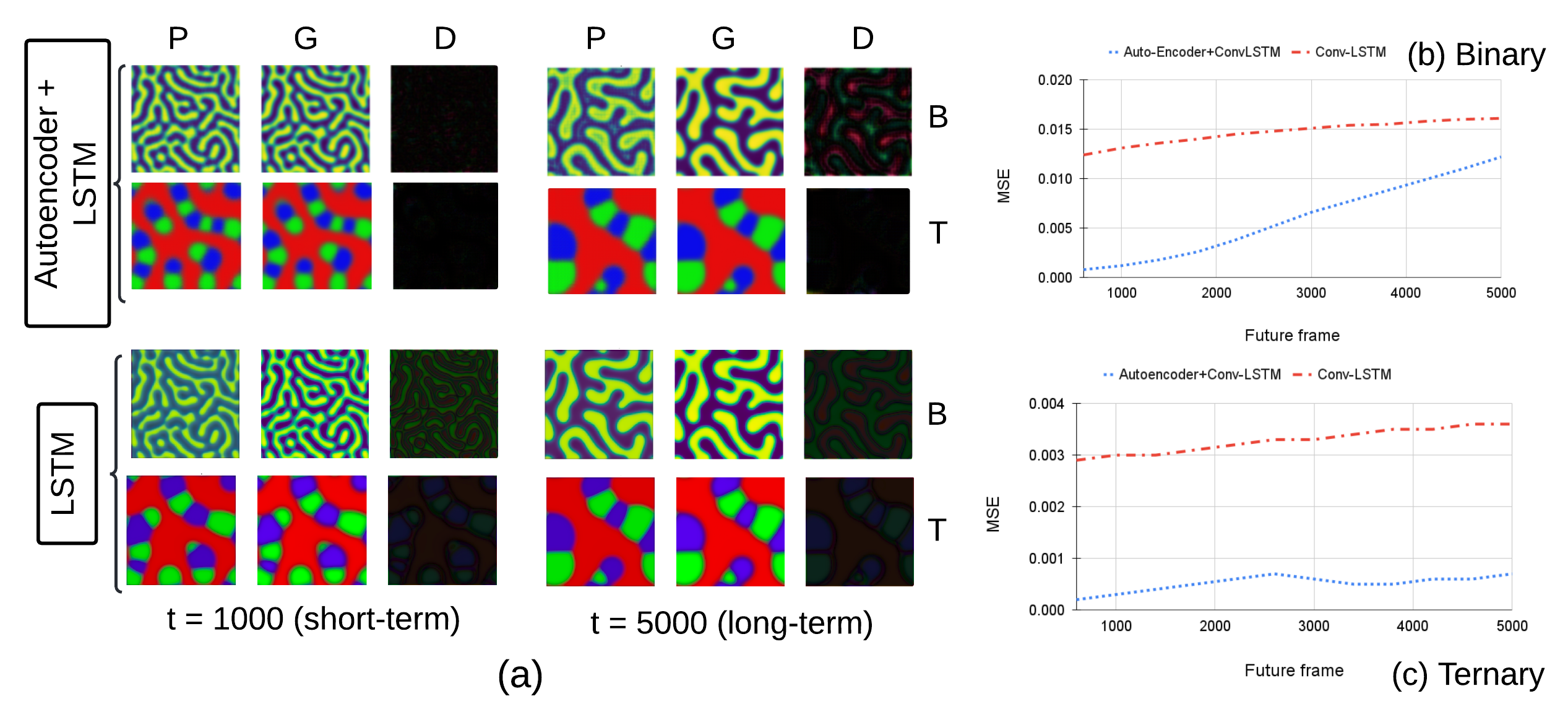}
\caption{(Color online) Here, P = predicted microstructure, G = ground truth or true microstructure, D = pixel-wise difference, B = binary, T = ternary. (a) We conduct our data-driven study with and without the encoder-decoder network (in Fig.~\ref{fig_schematic}). We show the prediction for the next frame in sequence ($t+10$) based on the input sequence of two-frame clips (\textit{i.e.}, frames $t-10$ and $t$). Microstructural differences are visible when comparing the predictions made by both the networks in the ternary system, with convolutional autoencoder+LSTM showing negligible accuracy loss in the heat map. (b, c) The MSE calculated between the predicted and true microstructures in both binary and ternary systems remains low when the autoencoder is used in the surrogate model. This signifies that the autoencoder efficiently reduces dimension representation of the input frames, which are then processed more efficiently by RNN for forecasting. We use LSTM for reference. We use colors for visualization of binary (blue: $A$ and yellow: $B$) and ternary (green: $A$, blue: $B$, and red: $C$) microstructures.}\label{fig_encoder}
\end{figure}

Next, we use solely the convolutional autoencoder and a combination of convolutional autoencoder and LSTM to predict future frames in the same sequence. The surrogate model takes a sequence of two-frame clips (\textit{i.e.}, frames $t-10$ and $t$) to forecast a single output at time $t+10$. We find autoencoder+LSTM performs better than LSTM for future predictions when we compare the heat map of the ternary microstructures both at short and long times (Fig.~\ref{fig_encoder}a). This is further quantified by the estimated MSE that increases monotonically over time series forecasting, with the autoencoder+LSTM gives lower MSE (Figs.~\ref{fig_encoder}b,~\ref{fig_encoder}c). These observations demonstrate that the input frames with compact representation lead to more accurate time series forecasting by RNN models. Therefore, we train the RNN models using the encoder-generated frames for future predictions. Based on the input protocol, to predict frame $t = 1000$, input frames $t \in \lbrace980,\ 990\rbrace$ are used. Thus, relative to the last timestep from the training dataset (\textit{i.e.}, $t = 900$), $t = 1000$ and $t = 5000$ represent short- and long-term predictions, respectively. Note that the predictions are made from the test dataset, which the model never ``sees'' during training.

\begin{figure}[ht]
\centering
\includegraphics[width=0.5\linewidth]{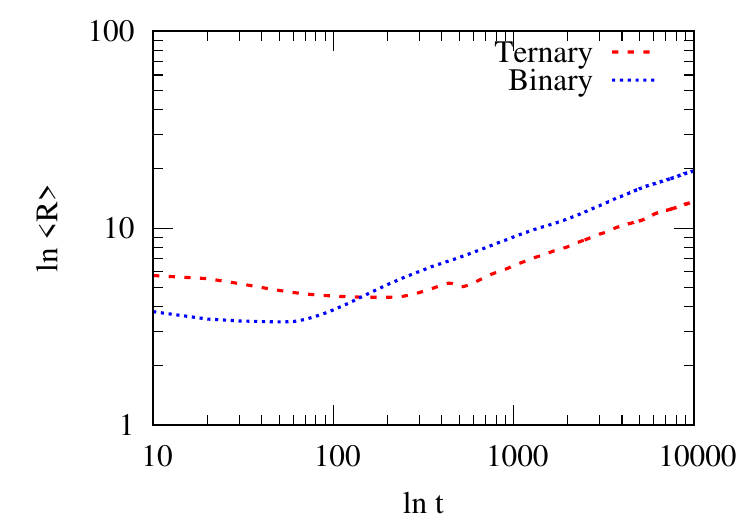}
\caption{Typical evolution of the average domain size ($\left\langle R(t) \right\rangle$) resulting in binary and ternary systems is shown.}\label{fig_domain_size}
\end{figure}

\begin{figure}[htbp]
\centering
\includegraphics[width=0.48\textwidth]{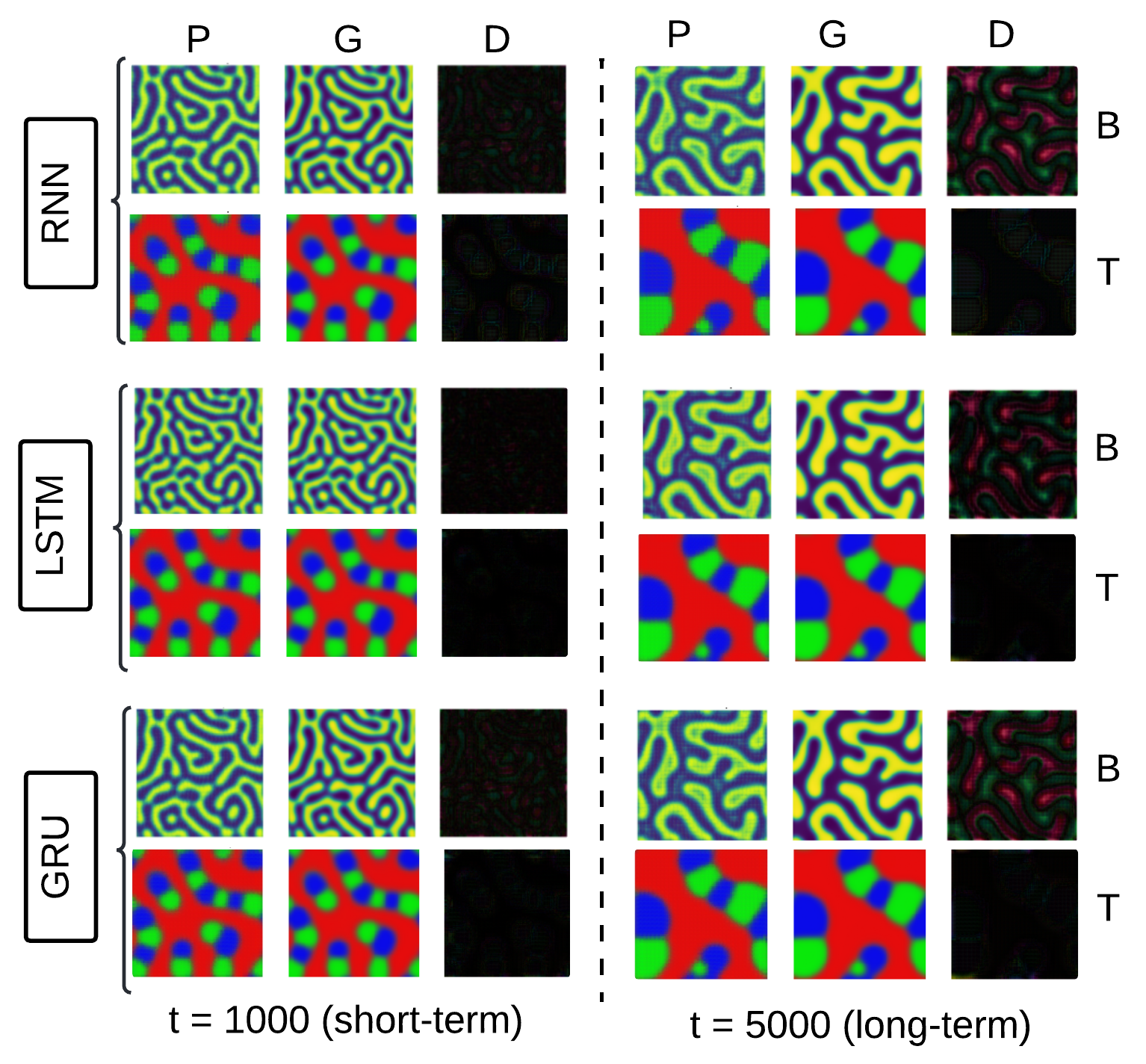}
\caption{(Color online) Here, P = predicted microstructure, G = ground truth or true microstructure, D = pixel-wise difference, B = binary, T = ternary. The RNNs in the surrogate model predict the next frame in sequence ($t+10$) based on the input sequence of two-frame clips (\textit{i.e.}, frames $t-10$ and $t$). Compared to the pixel-wise differences between the ground truth and predictions, the binary system shows large errors, while the ternary system shows very small errors that are difficult to visualize. Among RNNs, simple RNN shows visible differences in the heat map in both binary and ternary systems and short and long times. The error is quantified in Fig.~\ref{fig_mse}, which demonstrates that LSTM produces the lowest MSE. We use colors for visualization of binary (blue: $A$ and yellow: $B$) and ternary (green: $A$, blue: $B$, and red: $C$) frames.}\label{fig_train_micro}
\end{figure}

\begin{figure}[htbp]
\centering
\subfloat[Binary mixture]{\label{a1}\includegraphics[width=0.48\textwidth]{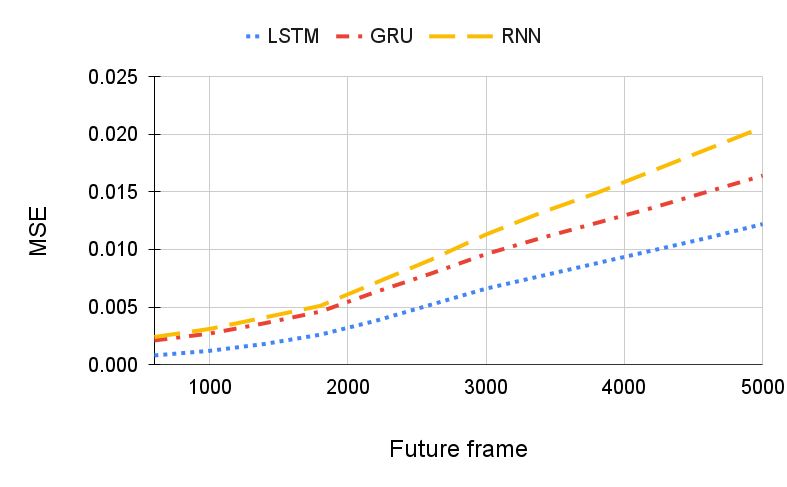}}
\subfloat[Ternary mixture]{\label{b1}\includegraphics[width=0.48\textwidth]{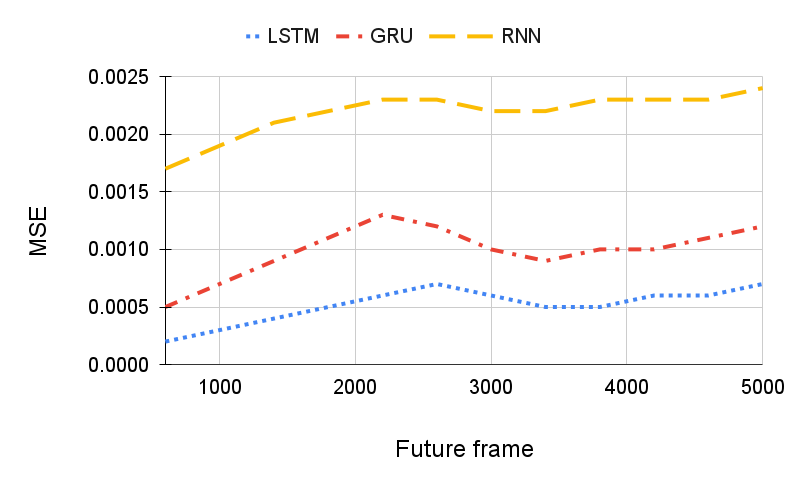}}
\caption{The results are obtained from temporal extrapolation tests for forecasting the next frame in sequence. We quantify the MSE between ground truth and predictions corresponding to Fig.~\ref{fig_train_micro}. The predictions are made on the testing dataset, $t \in \lbrace900,\ 5000\rbrace$, for (a) two-phase and (b) three-phase microstructure evolution. The trends in MSE signify that model accuracy increases in the following order: simple RNN < GRU < LSTM. Compared to the binary system, the lower MSE during three-phase evolution suggests the role of the ``third'' phase in slowing down the domain growth.}\label{fig_mse}
\end{figure}

In Fig.~\ref{fig_train_micro}, we show short-term and long-term forecasting of microstructure evolution using different RNNs. In the context of the morphology, the short-term predictions closely resemble the ground truth, while significant discrepancies appear in the long-term, which can be visualized in the heat map representing the point-wise difference from the ground truth. Since the differences between the predictions and ground truth accumulate over time series since the training data, visible differences appear in long-term predictions. The forecasting accuracy of different RNN models is shown in Fig.~\ref{fig_mse}. The lowest value of MSE associated with different RNNs signifies that their accuracy increases in the following order: simple RNN < GRU < LSTM. It is evident that LSTM yields better accuracy and long-term prediction ability for the microstructure evolution problem. Also, compared to the binary mixture (MSE < 0.01), the three-phase mixture shows very low MSE (< 0.001). This is likely due to the role of the ``third'' component in minimizing the symmetry-breaking effect in the ternary mixture, allowing for a much slower, steadier evolution of the microstructure over an extended period of time~\cite{ghosh_pccp,ghosh2020_jcp}. To quantify this, we study the average domain size ($\left\langle R \right\rangle$: a measure of characteristic length scale) of component $A$ in the microstructure by circularly-averaged structure function~\cite{zhu1999coarsening,saswata2020},
\begin{equation}\label{eq_sf}
S_{A}(k,t) = \frac{1}{N}\left\langle c_{A}^*(\textbf{k},t)c_{A}(\textbf{k},t)\right\rangle,
\end{equation}
where $N$ is the lattice size, $c_{A}(\bm{k},t)$ is the Fourier transform of $c_{A}(\bm{r},t)$, and $k$ is the magnitude of the wave vector $\bm{k}$. Thus, temporal evolution of $\left\langle R \right\rangle$ can be approximated by
\begin{equation}\label{eq_first_moment}
\left\langle R(t) \right\rangle = \frac{\sum S_{A}(k,t)}{\sum k S_{A}(k,t)}.
\end{equation}
The results are shown in Fig.~\ref{fig_domain_size}, which demonstrates the larger $\left\langle R(t) \right\rangle$ and different coarsening kinetics of the binary system compared to the ternary counterpart.

\begin{figure}[htbp]
\centering
\subfloat[Effect of frame interval]{\label{a2}\includegraphics[width=0.48\textwidth]{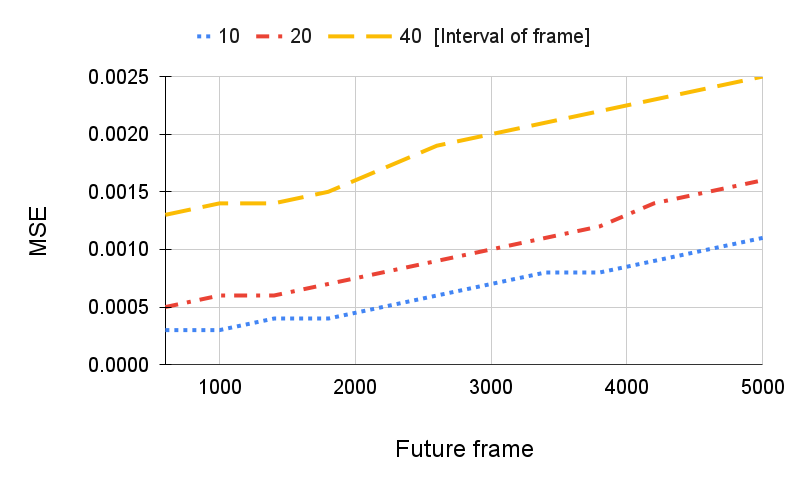}}
\subfloat[Effect of domain size]{\label{b2}\includegraphics[width=0.48\textwidth]{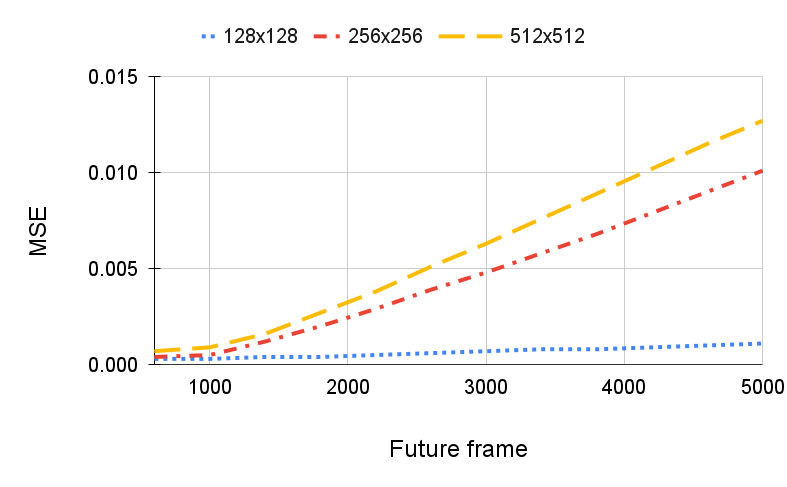}}
\caption{We investigate the effect of the (a) frame interval and (b) frame size on model performance. The results are obtained from temporal extrapolation tests for forecasting the next frame in sequence. (a) The MSE calculated between the predicted and true microstructures increases with the increasing interval ($i$) of the input two-frame clips (\textit{i.e.}, $t-i$ and $t$) to predict the next frame in sequence (\textit{i.e.}, $t+i$). We test with $i$ = 10, 20, and 40. (b) The MSE increases with the increasing domain size of the simulation box. This is attributed to finite-size effects during microstructure evolution, with the large simulation domain allowing for more convoluted, bi-continuous morphologies to form. In these studies, we have used the LSTM model and the ternary system for reference.}\label{fig_domain}
\end{figure}

We find that LSTM has more predictive power (lowest MSE) among the RNN variants. Thus, we next study the sensitivity of the prediction accuracy on the frame interval and frame size input to the surrogate model (Fig.~\ref{fig_schematic}). This is illustrated in Fig.~\ref{fig_domain}. The forecasting error increases significantly with the increase of both the frame interval and frame size, with the differences between the MSE curves widen over time series forecasting. As the deviation from the ground truth in spatial statistics increases with long-range forecasting, the test MSE increases considerably. Also, the error increases drastically with the increase of frame size (see 512 $\times$ 512 in Fig.~\ref{fig_domain}). This observation is attributed to the finite-size effects that arise during microstructure evolution in a large simulation domain, allowing for more randomness in generating interconnected bi-continuous morphologies. In the present study, we primarily use a frame interval of 10 and domain size of 128 $\times$ 128 for reference, the settings of which provide a good balance between computational accuracy (lowest MSE) and efficiency.

\subsection{Multi-frame ahead microstructure forecasting by RNNs}\label{sec_far}
\begin{figure}[htbp]
\centering
\includegraphics[width=0.6\textwidth]{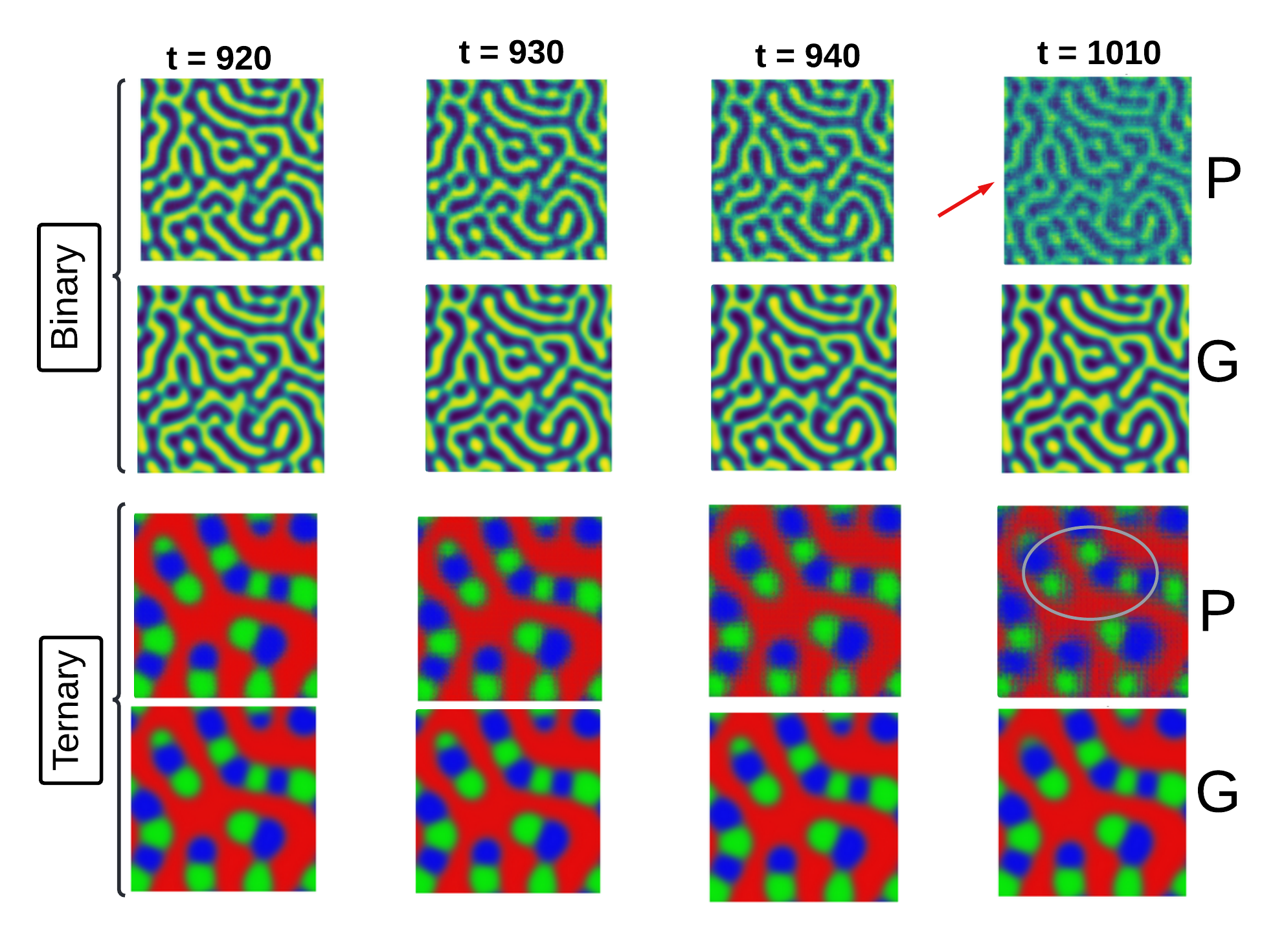}
\caption{(Color online) We perform temporal extrapolation tests for forecasting 20 frames (\textit{i.e.}, $t = 200$ timesteps) in the time series. After the initial input of frames $t-10$ and $t$ (\textit{i.e.}, ground truth) to predict frame $t+10$, the subsequent input consists of frames $t$ (true) and $t+10$ (predicted) to forecast frame $t+20$; then we input frames $t+10$ and $t+20$ (\textit{i.e.}, past predictions) to forecast frame $t+30$, and so on. Here we only show a few predictions in time series between $t = 920$ (1st frame) and $t = 1010$ (10th frame). The output becomes increasingly blurry due to forecasting error accumulation over the prediction time series. Also, the binary frames become more blurry than the ternary frames; for example, let us compare $t = 1010$ between them. The resulting MSE is quantified in Fig.~\ref{fig_future_micro}. Here, we apply LSTM as the RNN model. We use colors for visualization of binary (blue: $A$ and yellow: $B$) and ternary (green: $A$, blue: $B$, and red: $C$) microstructures.}\label{fig_future_micro}
\end{figure}

\begin{figure}[htbp]
\centering
\subfloat[Binary mixture]{\label{a3}\includegraphics[width=0.48\textwidth]{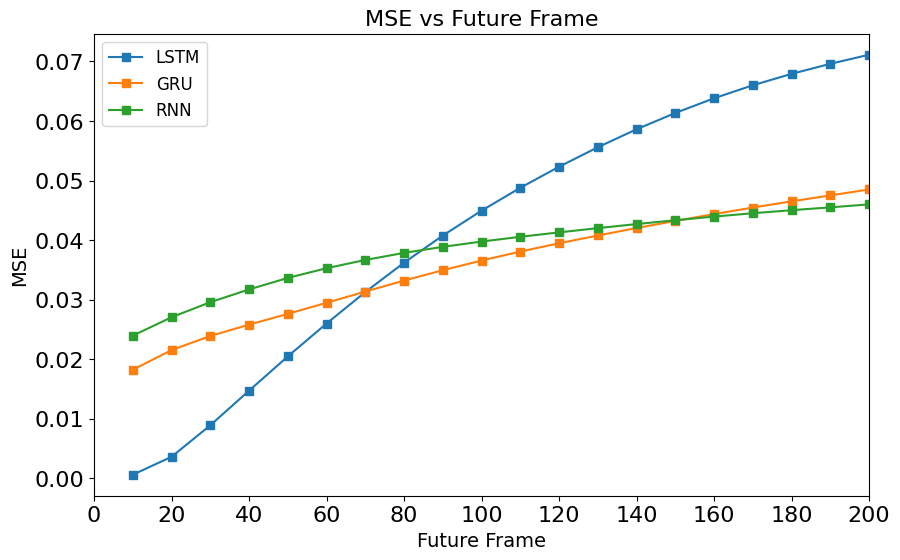}}
\subfloat[Ternary mixture]{\label{b3}\includegraphics[width=0.48\textwidth]{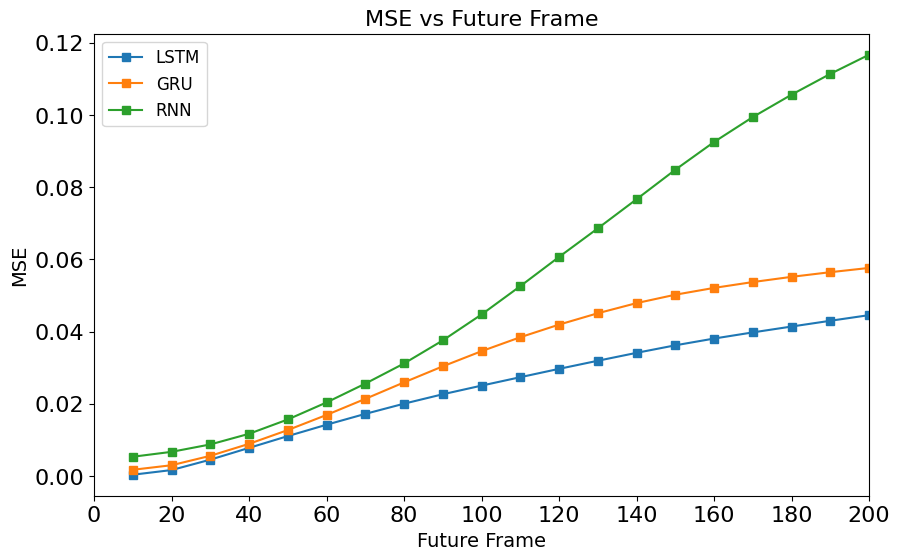}}
\caption{We quantify the MSE between ground truth and predictions corresponding to Fig.~\ref{fig_future_micro}. As detailed in Fig.~\ref{fig_future_micro}, we input true frames $t \in \lbrace900,\ 910\rbrace$ to predict frames in time series $t \in \lbrace920,\ 930, ...,1110\rbrace$. We calculate the MSE between each prediction and its ground truth while forecasting (a) binary and (b) ternary frames. The random noise that originates after each prediction accumulates over time series to amplify MSE in future predictions. On average, LSTM is the best model with the lowest MSE for forecasting microstructure evolution.}\label{fig_mse_future}
\end{figure}

 Until now, RNNs are used for temporal extrapolation tests for output one frame forward (Sec.~\ref{sec_next}), \textit{i.e.}, for an input of two consecutive frames ($t-10$ and $t$), predict the next frame in the sequence: $t+10$. How does the RNN model perform when forecasting multiple time frames in the future? In this case, the model's output is used again as the input for forecasting the next frame. We take the first prediction $t+10$ and include it in the next batch of input clips (\textit{i.e.}, $t$ and $t+10$); then we use the model again to predict the next frame at $t + 20$; and we repeat the same process until $t + 200$. Using this approach, we ask RNN to output 20 frames (or 200 timesteps) into the future since the last training data at $t = 900$. This is illustrated in Fig.~\ref{fig_future_micro}. The spatial resolution of the output frames progressively decreases when predictions are made increasingly into the future (see data at $t = 1010$). The image blurring and information loss in terms of the microstructure spatial features are clearly visible at $t = 1010$. The forecasting error is quantified by MSE in Fig.~\ref{fig_mse_future}. The MSE increases with the predictions increasingly made for the future. This is expected since the point-wise error from each prediction in the time series tends to accumulate, leading to a significant increase in MSE. When we compare different RNNs, their prediction accuracy on average increases in the following order: simple RNN < GRU < LSTM. Let us consider the ternary system. We observe a similar trend earlier in Fig.~\ref{fig_mse} with low MSE (< 0.001) when the RNNs are asked for one-frame ahead forecasting. However, high MSE (< 0.01) results in this case when forecasting is made for time series in the future based on past predictions instead of the ground truth.

\subsection{Integration of surrogate and phase-field model frameworks}
As illustrated in Secs.~\ref{sec_next} and~\ref{sec_far}, the time-evolved future frames predicted by the surrogate model are a good approximation to the corresponding phase-field-generated microstructures. Thus, replacing $n$ number of phase-field timesteps with $n$ number of surrogate model predictions can expedite the overall time evolution of the phase-field workflow. For example, the $n$th frame predicted by the surrogate model can be input to the phase-field solver as the initial configuration for simulating subsequent microstructure evolution. This approach reduces $n$ number of phase-field solve iterations from the solver without any appreciable loss in evolution trajectory. Such a hybrid approach of alternating between phase-field and surrogate models can save significant computing time and storage relative to the full-scale phase-field simulation~\cite{montes2021accelerating,oommen2022learning}.

\begin{figure}[htbp]
\centering
\includegraphics[width=0.48\textwidth]{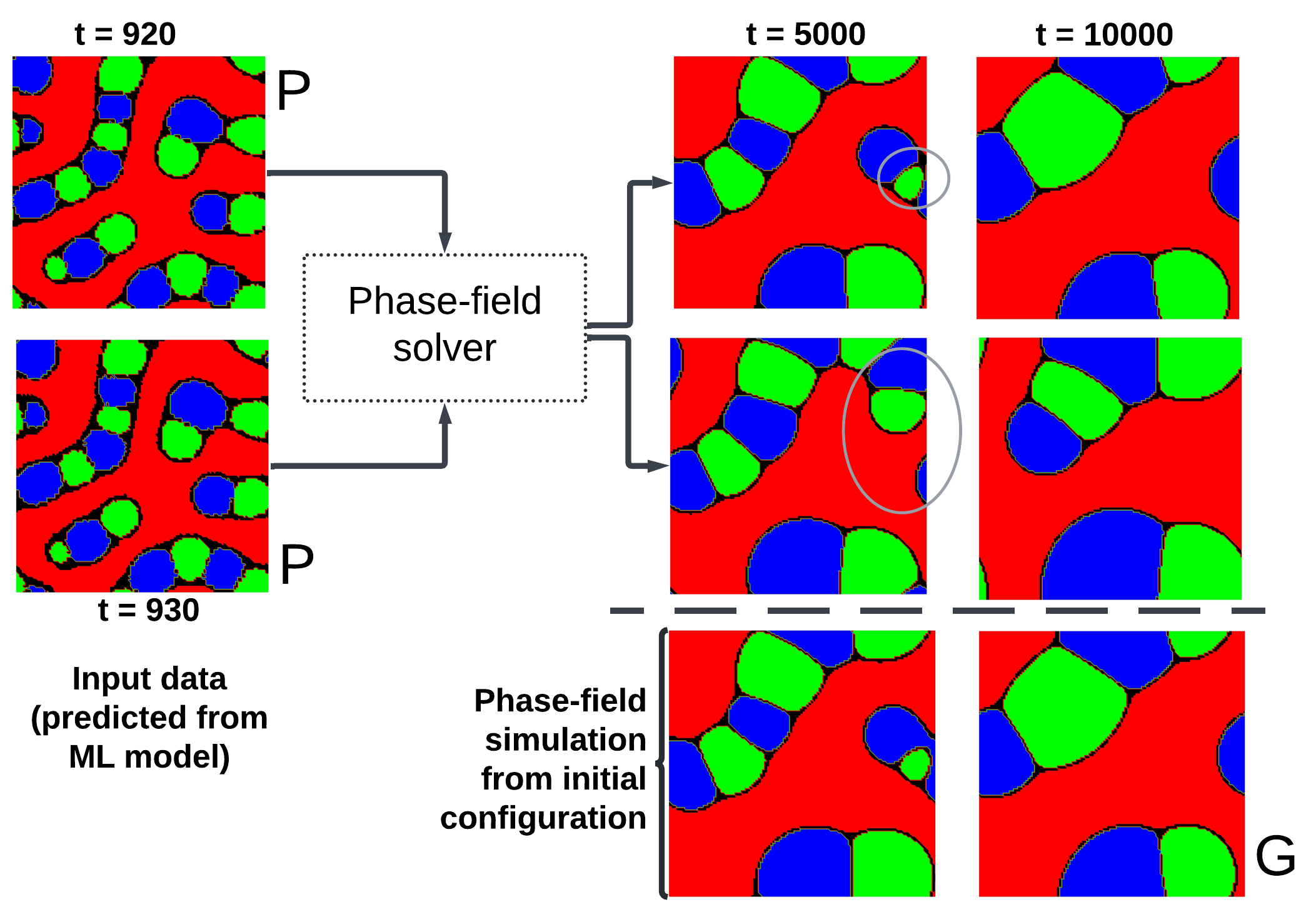}
\caption{(Color online) Here, P = predicted microstructure, G = ground truth or true microstructure. We use a ternary system for demonstration. From Fig.~\ref{fig_future_micro}, we use each prediction in time series as an initial configuration to the phase-field solver to simulate its long-time evolution. For the input of the first prediction $t= 920$, the phase-field solver produces a modified evolution trajectory at $t = 5000$; however, the correct/true evolution trajectory is reproduced at $t = 10000$. For the input of the next frame in sequence (\textit{i.e.}, $t= 930$), the evolution pathways deviate significantly from the ground truth at $t = 5000$ but produce reasonable output at $t = 10000$. The input of subsequent frames from the time series causes a runtime error in the solver, hence not shown here.  We use colors for visualization of binary (blue: $A$ and yellow: $B$) and ternary (green: $A$, blue: $B$, and red: $C$) microstructures.}\label{fig_solver}
\end{figure}

With proper training, the surrogate model takes a fraction of a second to forecast ten future frames, while the phase-field model takes about 20 seconds to simulate the same. Thus, integrating our data-driven approach with the phase-field solver makes an accelerated phase-field framework. In order to accelerate phase-field simulations, a given number of solver iterations can be replaced or ``leaped in time'' by the surrogate model predictions of corresponding timesteps. However, the question remains on how many phase-field steps can be replaced by the surrogate model to reproduce \textit{quantitative} microstructure evolution trajectory without loss in accuracy. To demonstrate this approach, we input the surrogate model predictions as an initial configuration to the phase-field solver to emulate microstructure evolution in the long-term. We note that the solver does not execute when the initial configuration is taken as frame $t+30$ and beyond, likely due to the considerable noise present in these frames that gets accumulated over time series forecasting. Thus, we can only simulate the evolution of the frames $t+10$ and $t+20$ in the phase-field solver (Fig.~\ref{fig_solver}). We compare their long-term evolution with the true evolution from the full-scale phase-field simulation implemented with a random initial configuration. Comparing the evolution trajectory at $t = 5000$, significant differences between the ground truth and predictions can be seen. However, these artifacts are eliminated in the long-term ($t = 10000$), leading to excellent agreement between them. 

\begin{figure}[htbp]
\centering
\includegraphics[width=0.48\textwidth]{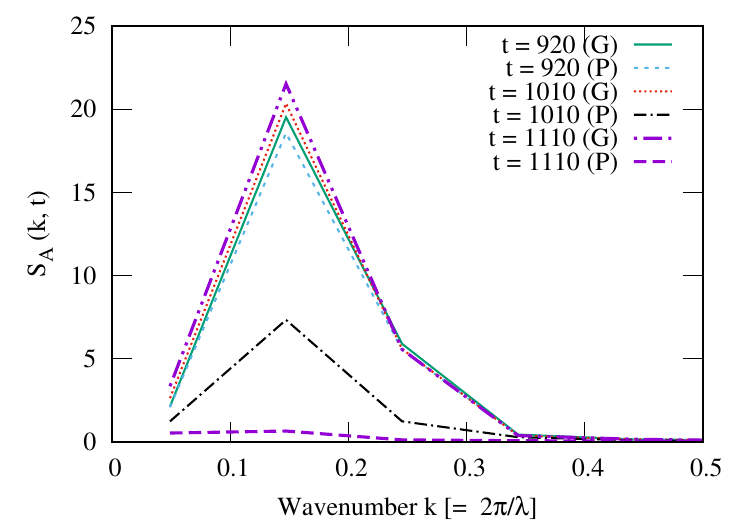}
\caption{Here, P = predicted (by LSTM), G = ground truth, $t$ is time, and $\lambda$ is a length scale. Fourier analysis of the structure factor distribution (of component $A$) in the predicted microstructure is compared to that in the true microstructure. When forecasting output for the entire time series (correspond to Fig.~\ref{fig_future_micro}), the signal of the power spectrum becomes increasingly weak as the microstructure progressively becomes blurred in future predictions.}\label{fig_sf}
\end{figure}

The above observations demonstrate that we can effectively replace 20 timesteps from the phase-field solver by the surrogate model predictions and then, using the prediction (say, $t+10$), run the phase-field simulation for the next 20 timesteps, and so on. Such alternate use of phase-field and surrogate models to generate the full microstructure evolution process can lower the computing cost by at least 40 times relative to the full-scale phase-field simulation, even with no loss in accuracy. To further quantify the merits of this hybrid approach, we use the structure factor analysis (Eq.~\eqref{eq_sf}) that accurately captures the statistical properties of microstructures, including the domain size. The $S_A$ in Eq.~\eqref{eq_sf} can capture statistically similar but quantitatively different local morphological features in $A$, with its maximum (say, $k_\text{max}$) signifies a measure of the average domain size. As illustrated in Fig.~\ref{fig_sf}, the domain size (given by $2\pi/k_\text{max}$) in frame $t = 920$ (\textit{i.e.}, $t+10$) agrees well with the ground truth. However, this agreement is lost for future time series predictions (\textit{e.g.}, $t = 1010$ and $t = 1120$), likely due to the accumulated noise over time series forecasting, weakening the signal of the power spectrum (\textit{i.e.}, low peak height at $k_\text{max}$). Such quantitative analysis of spatial statistics confirms that we can input frame $t = 920$ in the phase-field solver to produce the correct microstructure evolution trajectory without valuable information/accuracy loss on the salient physics of the phenomena of interest. We note that the MSE (Fig.~\ref{fig_mse_future}) with the predicted frame $t = 920$ is given by 0.002, which we can consider as \textit{benchmark} MSE for the ternary system to characterize the quality of the forecasting accuracy. Such analyses further demonstrate the good predictive capability of our robust yet simple data-driven framework.

\subsection{Transfer learning to forecast unseen composition and physics}\label{sec_new}
Our surrogate model shows satisfactory performance in predicting the time series of microstructure evolution in the ternary mixture ($A_{1/4}B_{1/4}C_{1/2}$) described earlier (Sec.~\ref{sec_next}). However, for a comprehensive understanding of the model behavior, it would be interesting to evaluate the predictive performance of the model on new microstructures generated using different material compositions or under different physical mechanisms. Thus, the question arises whether can we use the properly-trained surrogate model to forecast microstructures from different datasets/mixtures? This problem is equivalent to predicting future time series values of the unknown image sequence generated by different simulations. 

\begin{figure}[htbp]
    \centering
    \includegraphics[width=\textwidth]{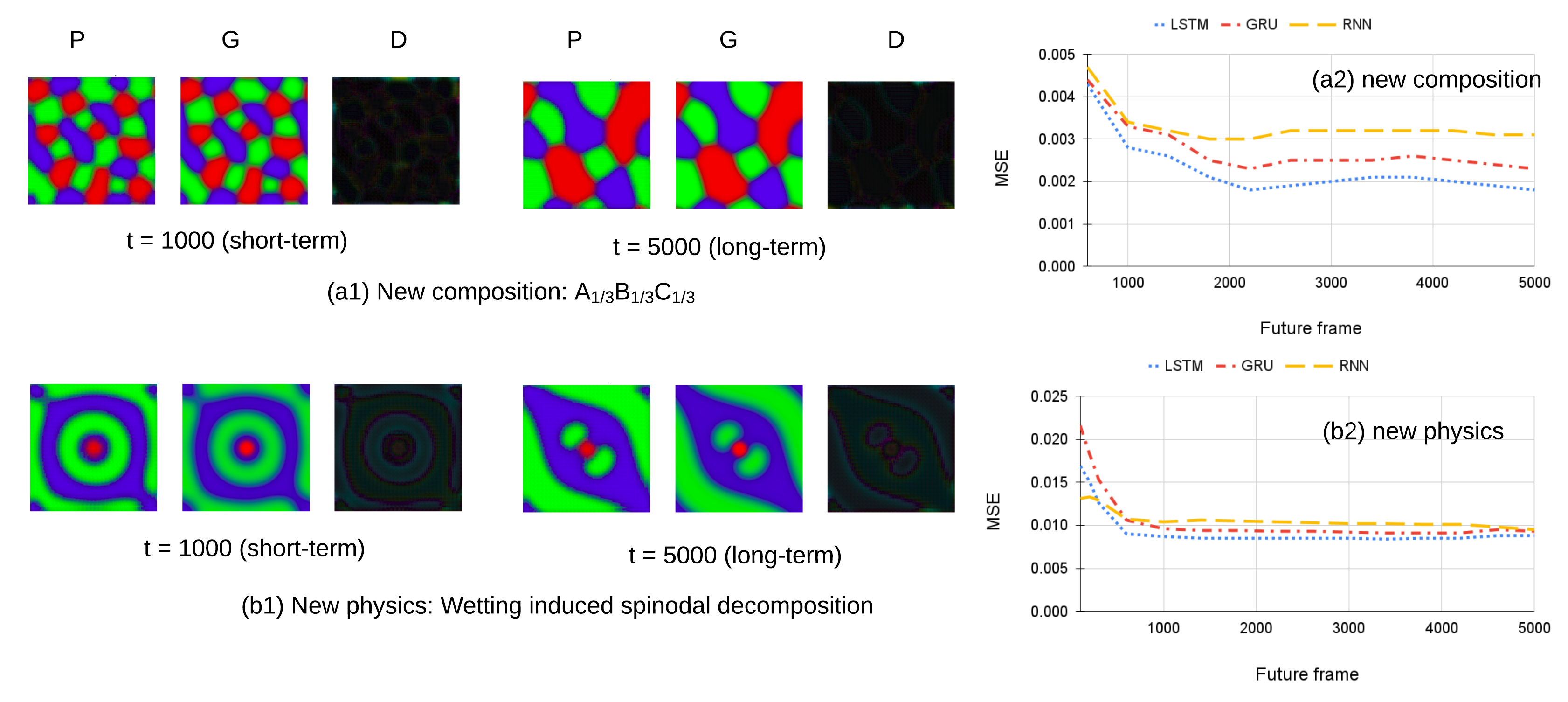}
    \caption{(Color online) Referring to Sec.~\ref{sec_new}, we use our pre-trained surrogate model for forecasting different time series of microstructure evolution produced with two different scenarios in mind: (a1) different material compositions without altering the spinodal decomposition physics known to the surrogate model, and (b1) altering parameters that lead to new physical mechanisms (\textit{e.g.}, wetting phenomenon) unknown to the model. We use the one-frame ahead forecasting (similar to Fig.~\ref{fig_train_micro}) as a baseline reference to investigate the performance of the surrogate model on unseen microstructures generated from different simulations. (a2) On average, the MSE remains low for predictions on unseen composition with no new physics involved. (b2) The MSE becomes high when new physics is present in the system for which evolution takes a completely different pathway from the model's knowledge. Typically, the early and late time trends in the MSE curves correspond to the rapid and slow regimes of microstructure evolution. We use colors for visualization of binary (blue: $A$ and yellow: $B$) and ternary (green: $A$, blue: $B$, and red: $C$) microstructures.}\label{fig_new}
\end{figure}

We use two different test datasets unknown to the trained surrogate model. In the first dataset, we vary the material composition ($A_{1/3}B_{1/3}C_{1/3}$ in Fig.~\ref{fig_new}a1) that leads to new microstructure morphologies, specifically the red-colored phase now takes a globular shape compared to its continuous morphology in Fig.~\ref{fig_pf_images}. However, in this case, the evolution principles remain the same: phase separation followed by the domain coarsening. In the second dataset, we vary material parameters ($\chi_{ij}$ and $\kappa_{i}$ in Eq.~\eqref{eq_ternary} following Ref.~\cite{ghosh2020_jcp}) to introduce surface-energy-driven phenomena (\textit{e.g.}, wetting) in the phase-separating system in addition to the domain coarsening effects. The wetting case is simulated after setting the thermodynamic interactions between the mixture components asymmetrically as $\chi_{AB}$ = 2.5, $\chi_{BC}$ = 3.5, $\chi_{AC}$ = 5.0; and $\kappa_A$ = 2.0, $\kappa_B$ = 6.0, and $\kappa_C$ = 6.0 (Sec.~\ref{sec_pf}). The wetting here refers to the selective preference of forming the low-energy interface between coexisting phases. This is illustrated in Fig.~\ref{fig_new}b1, where wetting-induced composition layers of the preferred phase develop surrounding the circular particle phase. Such ``ring pattern'' formation during spinodal decomposition was observed in experiments~\cite{aichmayer2003,karim1998}.

Next, we perform temporal extrapolation tests for forecasting the next frame in time series, similar to Fig.~\ref{fig_train_micro}. Good agreement (MSE < 0.005) between the ground truth and prediction can be seen when the new microstructure has a different morphology but follows the same evolution physics as the training dataset (Fig.~\ref{fig_new}a2). However, the MSE becomes high (> 0.01) when forecasting is made on new microstructures with unseen physics to the model (Fig.~\ref{fig_new}b2). In the MSE curves, the high MSE at the earlier times represents the fast evolution, which progressively becomes slower and steadier, approaching equilibrium at late times of phase separation. Compared to different RNN models, the prediction accuracy on average increases in the following order: simple RNN < GRU < LSTM. Such temporal extrapolation ability implies that irrespective of the microstructure obtained from different simulations with different thermodynamic and kinetic parameters, our pre-trained surrogate model at least \textit{qualitatively} understands the underlying general physical rules of evolution during spinodal decomposition of complex mixtures. To further improve forecasting accuracy in systems with unseen physics, the surrogate model needs to have some knowledge of the underlying physical evolution rules, more so in the ternary system, where more variations to the initial configuration are possible~\cite{ghosh2020_jcp,ghosh2024_fractal,saswata2020,amoabeng2017}.

\section{Discussion}\label{sec_discussion}
Phase-field simulations are used for quantitative microstructure evolution that accurately mimic experimental observations. Thus, phase-field datasets are considered as high-fidelity training data for developing predictive deep learning frameworks for forecasting a variety of microstructure evolution processes in diverse material systems. With this motivation, we develop the phase-field-informed surrogate model reported in this study that can be applied more generally to predict rich microstructure evolution problems due to spinodal decomposition, grain growth, order-disorder transformation, and alloy solidification, among others~\cite{Porter}. Undoubtedly, including experimental measurements within the training data will further enrich the classes of microstructure evolution.

As mentioned in Sec.~\ref{sec_intro}, spatiotemporal prediction of microstructure evolution using convolution RNNs has been studied considerably in the literature albeit for binary systems~\cite{oommen2022learning,ahmad2023accelerating,montes2021accelerating}. Also, the academic knowledge regarding the variance in efficacy when using different RNN variants in the surrogate model to predict the high-fidelity phase-field simulation is limited. It is well-established by these studies and others~\cite{farizhandi2023spatiotemporal,yang2021self,wu2023emulating,hu2022accelerating} that several phase-field simulation timesteps can be effectively replaced by deep model predictions in an alternating hybrid approach between these models, considerably accelerating the phase-field simulation framework. However, these studies do not demonstrate how precisely the predicted frames can be applied to generate microstructure evolution in the long term. Moreover, for a more complete picture, can we predict the evolution in a new dataset of ``unseen'' microstructures with new physics unknown to the trained model? To our knowledge, these questions have not been addressed by previous studies. Moreover, there has been no study on the applications of deep models for multiphase systems.

As a case study, we use multi-phase-field model-generated datasets of spinodally decomposing ternary microstructures for data analysis. Additionally, we present the typical academic example of binary spinodal commonly reported in the literature as a reference. The present study addresses three novel questions: How does the surrogate model predict the future frames in two-phase and three-phase spinodal systems? How can we quantitatively (\textit{e.g.}, through power spectrum analysis) determine the phase-field simulation steps that can be ``leaped in time” by deep model predictions for obtaining accurate long-term evolution? How accurately does our surrogate model predict the evolution history for unseen ternary microstructures simulated using new alloy composition and, particularly, physics?
 
\begin{figure}[ht]
\centering
\includegraphics[width=0.5\linewidth]{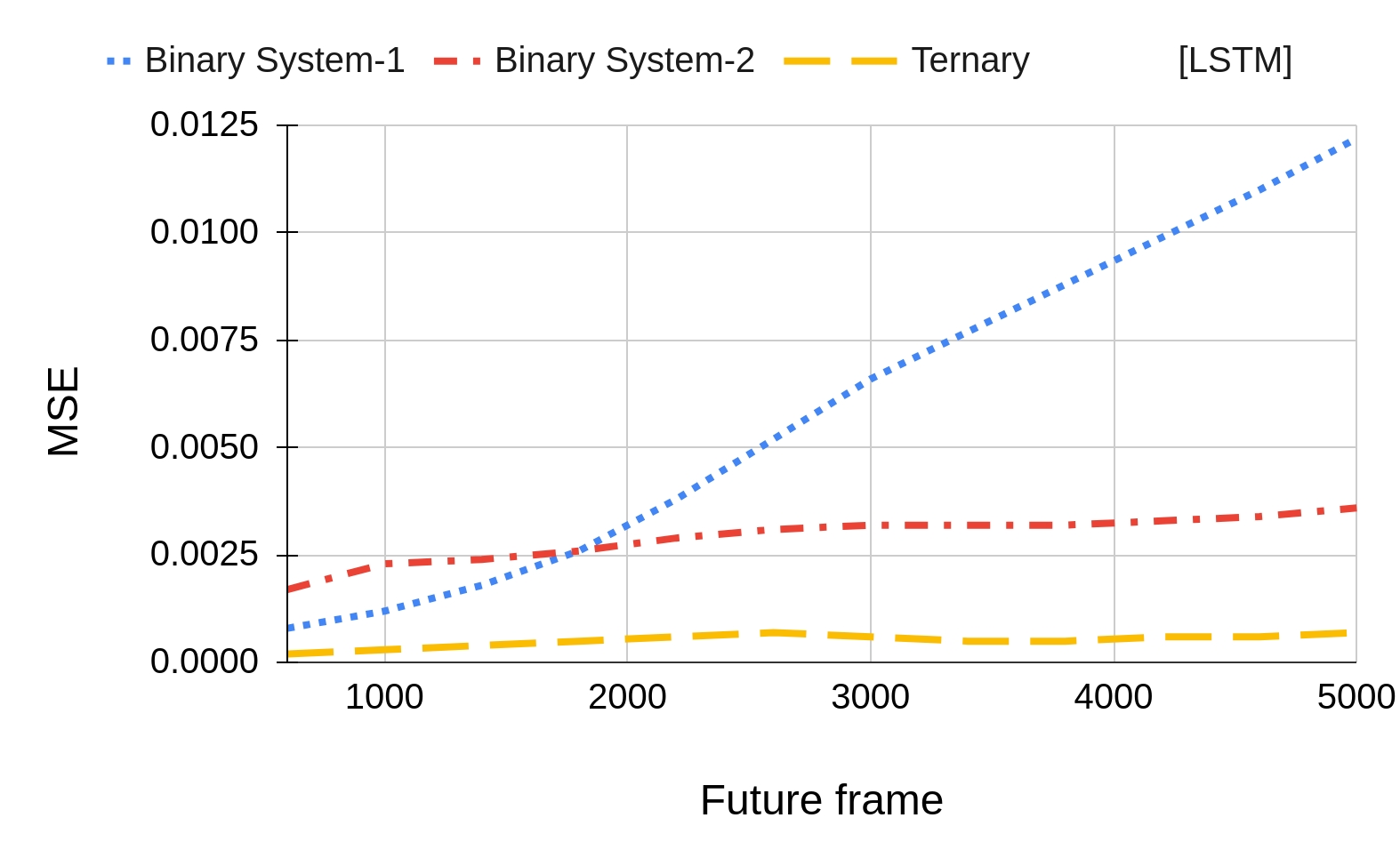}
\caption{We compare the prediction accuracy of the surrogate model between the ternary system and two different binary systems (or free energy functions). The binary system-1 is simulated following Eq.~\eqref{eq_binary}, and binary system-2 is implemented by approximating the ternary formulation in Eq.~\eqref{eq_ternary} to simulate a critical binary composition. For reference, we use LSTM for the next frame forecasting following Sec.~\ref{sec_next}.}\label{fig_comparison_free}
\end{figure}

We must emphasize that we have used different extensions of the free energy functions when comparing the binary (Eq.~\eqref{eq_binary}) and ternary systems (Eq.~\eqref{eq_ternary}), which may lead to qualitative differences. Unfortunately, the present ternary formulation can not be applied to simulate binary systems due to the constraints that need to be satisfied for which complete freezing of one of the components is not possible (see~\cite{ghosh2020_jcp} for details). However, we can approximate the phase-field model to work with a critical ($50:50$) binary composition but with a negligible amount of the third component; such preliminary analysis (Fig.~\ref{fig_comparison_free}) further confirms the general trends of lower MSE in predicting ternary systems, similar to the observations reported in Fig.~\ref{fig_mse}.

We acknowledge that the current implementation, while using a simple semi-implicit solver, serves as a reasonable first step for coupling surrogate models with traditional phase-field simulations. The focus here is on demonstrating how a surrogate model can effectively accelerate phase-field simulations, even with a basic solver utilizing a small time step. We clarify that the acceleration from surrogate models comes not just from a single-step improvement but from eliminating the need to compute multiple time steps explicitly in a phase-field solver, reducing the overall simulation time. The RNN learns the dynamics and predicts future states directly, bypassing many intermediate time steps. It is straightforward to understand that this acceleration effect will be manifold when working with large-scale systems to obtain long-term evolution history. Future work will explore combining the surrogate model with implicit or more advanced methods~\cite{ghosh2022tusas}, which could allow for larger time step size, reducing the number of simulation time steps needed without sacrificing numerical accuracy. This gradual improvement will further optimize the phase-field solver and surrogate model for speed and efficiency.

We must admit that we accelerated the numerical simulations efficiently using the available computing resources. For example, we simulated the phase-field simulations in parallel utilizing eight threads in an AMD Reyzen 5600 CPU operating at 4.4 GHz supported by 16 GB of RAM. We then moved the generated data to the GPU-accelerated Kaggle~\cite{kaggle} environment (NVIDIA T4 $\times$ 2) for data-driven analysis. It is to be noted that the readily parallelizable nature of convolutional operation makes convolutional CNNs and RNNs more suitable for implementation on GPUs~\cite{pandey2022transformational}. We emphasize that our surrogate model needs to be trained once before it can be used to predict future frames without further retraining. On average, the training time was $\approx$124 s, while the inference time for predicting a future frame using the trained model was $\approx$0.8 s. Undoubtedly, when more CPU/GPU resources become available, these times will be further reduced. Moreover, for the given ternary spinodal problem, we have found that surrogate predictions can safely replace 20 phase-field steps without sacrificing the prediction accuracy regarding the long-term microstructure evolution. For a simulation domain size of $512 \times 512$, the phase-field calculation typically takes 40 s to solve 20 timesteps, while the surrogate model takes $\approx$0.8 s to predict the same number of future frames, leading to a speedup ratio of 50 (excluding the training time). This gain in computing time will accumulate when alternate phase-field and surrogate solvers will be used to expedite the entire simulation workflow. With increasing domain size, the speedup in computing time will become even more evident. We acknowledge that, among others, using more advanced numerical solution methods in the phase-field simulations and the type and number of threads in parallel computing framework will affect the computing cost and, hence, the speedup ratio. We will contemplate these problems in the future.

Moving forward, we can explore several improvements to enhance the accuracy, predictability, and acceleration performance of our data-driven approach. These modifications fall into two broad categories: developing the surrogate model and its behavior to the complexity of microstructure evolution.

We use an autoencoder as a reference for image compression so that the forecasting ability of the RNN models can be compared efficiently. Given the non-linearity of the microstructure evolution problem, alternate dimensional reduction techniques such as isometric feature mapping~\cite{hu2022accelerating}, uniform manifold approximation~\cite{hu2022accelerating}, kernel principal component analysis~\cite{williams2014kernel}, diffusion maps~\cite{williams2014kernel}, and dynamic mode decomposition~\cite{kutz2016dynamic} can be explored. Similarly, alternative RNN options such as Eidetic-RNN~\cite{yang2021self}, 3D CNNs~\cite{3dcnn}, PredRNN~\cite{predrnn}, generative adversarial networks~\cite{geron2022hands}, and transformer models~\cite{geron2022hands} can be explored for improved performance in image sequence prediction in both short and long times. 

Typical deep models do not incorporate physical rules (e.g., partial differential equations) of the spatiotemporal process. Thus, RNNs can capture ``new'' microstructure patterns, possibly related to unknown physics, purely through data-driven learning. Although RNNs are not explicitly informed by governing equations, they can learn complex patterns and temporal dependencies from large datasets, effectively approximating unknown physical phenomena, which may or may not fully align with true underlying physics. In this context, physics-informed deep models such as physics-informed neural networks~\cite{ren2022phycrnet} and physics-informed Gaussian processes~\cite{zhong2024label} explicitly incorporate known governing equations (\textit{e.g.}, continuity, conservation of mass) into the training process, ensuring that the predictions follow the fundamental physical laws. Moreover, deep learning models are essentially ``black box'' models with no guidance on the measurement uncertainty and on the input parameters essential for forecasting. Thus, a fusion of deep models with classical statistical models such as variational Bayesian inference, Monte Carlo dropout, and mixture density networks could be beneficial~\cite{wikle2023statistical,abdar2021review}. Another common issue with the deep models, interpretability, can be improved, for example, using non-negative matrix factorization that can decompose the microstructure data into additive parts, which might correspond to meaningful physical components of the microstructure~\cite{lee1999learning,lee2000algorithms}.

Although we have used a basic cell structure in RNN models, the resultant forecasting MSE remains low (< 0.001). However, the best solution will depend on other parameters such as number of frames in the input sequence, the time interval between frames, the number of frames/sequence used for model training, the number of convolutional layers, the number of memory cells in RNN, the activation functions in the neural nets, etc.~\cite{geron2022hands,montes2021accelerating}, which can be tuned for further improvement of our model performance. However, varying so many parameters will be impractical, given our aim is to explore the forecasting ability of the data-driven framework for a fixed optimal set of parameters/constraints. Moreover, to make our model more robust for diverse applications, particularly with scarce datasets, the transfer learning method, which uses a pre-trained network with completely different training dataset from material microstructures, such as ResNet~\cite{resnet}, EfficientNet~\cite{efficientnet} could provide more efficient prediction of microstructure evolution after proper fine tuning~\cite{farizhandi2022deep}. 

 Moreover, we can use our framework to study microstructure statistics requiring a design-of-experiments approach~\cite{ghosh2019uncertainty,ghosh2020statistical} in which the model and material parameters vary within predefined ranges to address the increasing data demand to predict microstructure evolution problems for unknown values of the input configuration such as alloy composition not directly ``seen'' by the model. It would also be interesting to use our framework to predict microstructure evolution in 3D since the diffusional processes and resultant length scale of the microstructures are very different in 3D compared to 2D even when all other parameters remain the same~\cite{ghosh20183d}. Finally, our framework will be increasingly improved based on the above recommendations, enabling it to handle more complexity in material microstructures with stronger spatiotemporal correlations between them along their evolution trajectory.   

\section{Conclusions and outlook}\label{sec_summary}
We use a phase-field-informed data-driven framework combining convolutional autoencoder and RNN architectures for rapid, accurate estimation of microstructure evolution in binary and ternary mixtures. We use different variants of RNN, including simple RNN, LSTM, and GRU, to compare their efficacy in developing surrogate deep-learning models for phase-field predictions in the context of a time series problem. We highlight the following observations from this work:

\begin{itemize}
\item On average, the results obtained by phase-field simulations and predicted by the surrogate model are in good agreement, demonstrating the learning ability and predictive power of our workflow. While the short-term predictions closely resemble the ground truth, error accumulates over time series forecasting. Thus, visible differences appear in long-term predictions, particularly in the high-gradient interfacial regions of the microstructure morphology (Figs.~\ref{fig_train_micro},~\ref{fig_future_micro}). We observe a satisfactory performance (low MSE < 0.001) over the time series of microstructure forecasting.

\item We have shown that RNN models can correctly predict the short-term spatiotemporal dynamics of microstructure evolution, but their forecasting ability decreases in the long-term. Also, the MSE calculated between ground truth and predictions with the RNN models increasingly deviates from each other over time. Among all RNN models, LSTM demonstrates the best learning ability, prediction accuracy, and long-term forecasting ability with the lowest MSE due to its ability to retain long-term dependencies in data due to the specialized gating mechanisms (input, forget, and output gates) to manage the flow of information. Other RNN variants, such as simple RNN and GRU, are expected to perform similarly when modeling short sequences, small datasets, or data with minimal noise, where long-term dependencies are not crucial. However, these effects have not been tested presently in this study.

\item Despite the satisfactory performance of all RNN models (\textit{i.e.}, low MSE), their learning rate and predictive power vary with the nature of the microstructure evolution problem to which they are applied. The agreement between the predicted and true microstructures is better in the ternary system than in the binary system. These observations can be attributed to the time evolution of the microstructure spatial statistics with drastic topological changes, including inherent randomness during binary phase separation, and the role of the ``third'' element in the ternary system that may provide a barrier to long-range diffusional transport processes, leading to a more slow, steadier microstructure evolution compared to the binary systems (Fig.~\ref{fig_domain_size}).

\item With proper training, our surrogate model makes time series forecasting in a fraction of a second with increasing loss in accuracy over time. Using temporal extrapolation tests for multi-frame ahead forecasting (Fig.~\ref{fig_future_micro}) and the Fourier analysis of the predictions (Fig.~\ref{fig_sf}), we demonstrate how many time series values can be used as an initial configuration to the phase-field solver for generating correct microstructure evolution trajectory. Consequently, these predictions can effectively replace the corresponding timesteps in the phase-field solver, which takes significant time to compute. In this way, the surrogate-model assisted phase-field sovler can significantly accelerate the microstructure evolution workflow with negligible loss in accuracy. 

\item We have tested the general ability of our surrogate model to forecast new microstructure evolution processes unknown to the model (\textit{i.e.}, to predict future values of different image sequences). We use an approach similar to transfer learning in which the surrogate model is trained with a known ternary dataset (Fig.~\ref{fig_pf_images}) and subsequently applied to predict new microstructures created with different material compositions or underlying evolution physics. The RNN models, particularly LSTM, show quantitative prediction accuracy (MSE < 0.005) on new microstructures with unknown composition (Fig.~\ref{fig_new}). However, the agreement is qualitative (MSE < 0.01) when the model forecasts evolution on new microstructures with unknown physics. Qualitatively speaking, these examples imply our model's intuition into the underlying physics and alloy chemistry of the microstructure formation, making it a useful general-purpose alternative to physics-based, PDE-based approaches to microstructure evolution. 
\end{itemize}
%different from those used for training

%----------------------------------------------------------------------------------------

% If you have acknowledgments, this puts in the proper section head.
\section*{Acknowledgments}
S. Ghosh acknowledges the support of the Faculty Initiation Grant from the Sponsored Research \& Industrial Consultancy Office, Indian Institute of Technology Roorkee, and Science and Engineering Research Board, Government of India.

\section*{Conflict of Interest Statement}
The authors have no conflicts to disclose.

\section*{Data Availability Statement}
The data that supports the findings of the study are available from the corresponding author upon reasonable request.

% Create the reference section using BibTeX:
%\section*{References}
%\bibliography{spinodal}

\end{document}